\documentclass[journal]{IEEEtran}

\usepackage{graphicx}
\usepackage{subfigure}
\usepackage{cite}

\begin{document}

\newcommand{\simgt}{\lower.5ex\hbox{$\; \buildrel > \over \sim \;$}}
\newcommand{\simlt}{\lower.5ex\hbox{$\; \buildrel < \over \sim \;$}}
\newcommand{\ohm}{$\Omega$ }
\newcommand{\del}{\partial}
\newcommand{\be}{\begin{equation}}
\newcommand{\ee}{\end{equation}}

\title{Gain Stabilization of a Submillimeter SIS Heterodyne Receiver}

\author{James~Battat, Raymond~Blundell, Todd~R.~Hunter, Robert~Kimberk, Patrick~S.~Leiker, and Cheuk-yu~Edward~Tong \IEEEmembership{Member, IEEE}%
\thanks{The authors are with the Harvard-Smithsonian Center for Astrophysics, Cambridge, MA 02138, USA (e-mail: jbattat@cfa.harvard.edu)}
\thanks{J. Battat was supported by the National Defense Science and Engineering Graduate Student Research Fellowship}
}

\markboth{IEEE Transactions on Microwave Theory and Techniques,~Vol.~53, No.~1,~January~2005}{Battat \MakeLowercase{\textit{et al.}}}

\maketitle

\begin{abstract}

We have designed a system to stabilize the gain of a submillimeter heterodyne receiver against thermal fluctuations of the mixing element.  
In the most sensitive heterodyne receivers, the mixer is usually cooled to 4 K using a closed-cycle cryocooler, which can introduce $\sim$1\% fluctuations in the physical temperature of the receiver components.
We compensate for the resulting mixer conversion gain fluctuations by monitoring the physical temperature of the mixer and adjusting the gain of the intermediate frequency (IF) amplifier that immediately follows the mixer.

This IF power stabilization scheme, developed for use at the Submillimeter Array (SMA), a submillimeter interferometer telescope on Mauna Kea in Hawaii, routinely achieves a receiver gain stability of 1 part in 6,000 (rms to mean).  This is an order of magnitude improvement over the typical uncorrected stability of 1 part in a few hundred.  
Our gain stabilization scheme is a useful addition to SIS heterodyne receivers that are cooled using closed-cycle cryocoolers in which the 4 K temperature fluctuations tend to be the leading cause of IF power fluctuations.  

\end{abstract}

%
\IEEEpeerreviewmaketitle

\section{Introduction}

\PARstart{S}{uperconductor} insulator superconductor (SIS) receivers are in use on many millimeter and submillimeter telescopes because of their good spectral line sensitivity.  
Their continuum sensitivity, however, does not usually reach the theoretical limit because of receiver gain fluctuations.  These arise predominantly from changes in mixer conversion gain, which result from physical temperature changes and local oscillator (LO) power changes.
Here we describe a technique to reduce the impact of physical temperature changes on mixer conversion gain variations by changing the intermediate frequency (IF) gain of the receiver in proportion to 
fluctuations of the physical temperature of the SIS junction used as a mixing element.




A number of radio observatories make use of liquid Helium filled cryostats so that temperature induced receiver gain fluctuations are minimized.  However, due to their lower degree of maintenance and upkeep, closed-cycle Helium cryocooler systems may be preferred.
For example multi-receiver systems with high heat loads, 
such as those in use at the 8-element Submillimeter Array (SMA)\footnote{The Submillimeter Array is a joint project between the Smithsonian Astrophysical Observatory and the Academia Sinica Institute of Astronomy and Astrophysics, and is funded by the Smithsonian Institution and the Academia Sinica.} \cite{refs:SMA} or the upcoming 64-element Atacama Large Millimeter Array (ALMA) \cite{refs:Wootten}, 
are dependent on the use of cryocoolers.
For these applications it is necessary to develop techniques to reconcile the need for highly stable receivers with the practical benefits of cryocoolers.  An obvious approach is to design more stable cryocoolers, another is to compensate for the fluctuations in the mixer's conversion gain.

In the Berkeley-Illinois-Maryland-Association Millimeter Array (BIMA) a heating resistor on the cold stage is used to compensate for cryocooler temperature fluctuations \cite{refs:plambeck}.  This reduces long time scale temperature drifts, giving 1 mK rms temperature stability over several hours when the data are averaged in 2 second blocks.  The cycling of the cryocooler's displacer also introduces temperature fluctuations on 1-2 second time scales, which are not treated by this technique.

To provide a large thermal capacity that will damp the fluctuations in the temperature of the cryocooler, it is possible to add a liquid helium reservoir.  Using this technique, a ten fold improvement in temperature stability, from 200 mK to 20 mK (peak to peak), has been demonstrated \cite{refs:sekimoto}.  The ALMA cryocooler also incorporates a helium reservoir and has achieved an rms temperature stability of 1 mK
at 4 K \cite{refs:anna}.

We take a different approach to receiver stabilization.  While monitoring the physical temperature of the mixer we actively vary the IF gain of the receiver to compensate for mixer conversion gain variations.  Such a scheme is straightforward to implement and can be applied to existing systems without redesigning the cryocooler.

We have developed and tested hardware and accompanying software to adjust the third stage gain of the low noise HEMT amplifier immediately following the mixer in proportion to the fluctuations in the physical temperature of the mixer.  We achieve nearly an order of magnitude improvement in receiver stability from 1 part in $\sim$800 to 1 part in 6,000 (rms to mean ratio) using a 33 millisecond integration time over 10 minutes.  This level of receiver stability equals that which would be obtained if rms fluctuations of the physical temperature of the mixer were held to $<$ 2 mK.  



\section{Theoretical Considerations}
The conversion gain of an SIS mixer has a negative temperature coefficient \cite{refs:Baryshev,refs:Kooi}.  
In simple terms, as the physical temperature, $T_{phy}$, of the mixer increases, the leakage current increases and the current-voltage relationship near the gap voltage becomes less sharp causing the mixer conversion gain, $G$, to decrease.
For small temperature changes we may write 
\be
	\frac{\Delta G}{G} \propto - \frac{\Delta T}{T_{phy}}
\ee
where $\Delta G$ and $\Delta T$ represent small fluctuations in the conversion gain and physical temperature of the mixer, respectively, and the minus sign reflects the anti-correlation between physical temperature and conversion gain.

At the SMA a two stage Gifford-McMahon (G-M) and a single Joule-Thomson (J-T) stage closed-cycle cryocooler maintain the SIS mixer at a nominal operating temperature of 4.2 K.  In our system, cryocooler-induced temperature swings of about 1\% (10-60 mK) result in mixer gain fluctuations of about 0.5\%.  Under the assumption that the $1/\sqrt{B\tau}$ fundamental gaussian radiometer noise and the instrumental gain fluctuations are statistically independent, \cite{refs:Tiuri} shows that they contribute to the fractional error in the system temperature by adding in quadrature to give:
\begin{equation}
\frac{\Delta T}{T_{sys}} = \sqrt{\left(\frac{1}{\sqrt{B\tau}}\right)^2 + 
	\left(\frac{\Delta A}{A}\right)^2}
\end{equation}
where $\Delta T / T_{sys}$ is the fractional error in the measured system noise temperature, $B$ and $\tau$ are the IF bandwidth and integration time, respectively, and $\Delta A/ A$ is the fractional gain fluctuation in the entire receiver chain.

The SMA receivers have a 2.5 GHz IF bandwidth and the digital correlator has a minimum integration time of 1 second, so the fundamental radiometer noise is $\simlt$ 2 x $10^{-5}$.  Therefore fractional instrumental gain fluctuations, typically of order $10^{-2}$-$10^{-3}$, dominate the system temperature fluctuations.  
In order to approach the fundamental noise limit of the receiver, the instrumental gain stability must be significantly improved.  

\begin{figure}
\centering
\includegraphics[angle=0, scale=0.42]{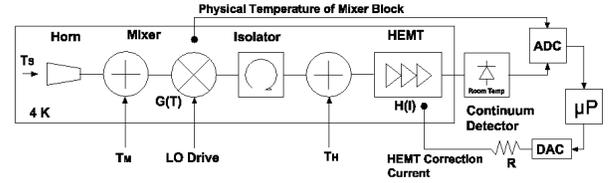}
\caption{A schematic of the SIS receiver and gain stabilization servo system.  The horn, mixer, isolator and HEMT amplifier are maintained at cryogenic temperatures by a cryocooler.  The physical temperature of the mixer, $T_{phy}$, is monitored by a diode thermometer.  The continuum detector measures the IF power.  An ADC records the mixer temperature and the detected continuum power simultaneously.  The microprocessor ($\mu$P) directs the DAC to generate a voltage, $V_{DAC}$, in direct proportion to measured mixer temperature fluctuations (see Equation \ref{servoEqn}).  This voltage is converted to a correction bias current for the third stage of the HEMT IF amplifier by the resistor, $R$.  The symbols, $G$, $H$, $T_S$, $T_M$ and $T_H$ are described in the text after Equation \ref{pif}.  }
\label{servoSchematic}
\end{figure}

Ignoring losses at the receiver input, the power output of an SIS receiver, shown schematically in Figure \ref{servoSchematic}, can be written in the Rayleigh-Jeans limit ($h\nu << kT$) as
\begin{equation}
\label{pif}
P_{IF} = k_BBGH\left(T_S+T_M+\frac{T_H}{G}\right)
\end{equation}
where 
\begin{eqnarray*}
k_B &=& \mbox{Boltzmann's constant}\\
B &=& \mbox{IF bandwidth}\\
G &=& \mbox{Mixer conversion gain}\\
H &=& \mbox{HEMT amplifier gain}\\
T_S &=& \mbox{Source temperature}\\
T_M &=& \mbox{Mixer noise temperature}\\
T_H &=& \mbox{HEMT amplifier noise temperature}
\end{eqnarray*}

If we assume that mixer conversion gain fluctuations are the dominant source of receiver gain fluctuations, and that $T_M$ and $T_H$ are independent of the physical temperature of the mixer and IF amplifier, then we can compute the necessary change in IF gain required to maintain a constant IF power from the following condition
\begin{equation}
\label{dpif}
\Delta P_{IF} = \frac{\del P_{IF}}{\del G}\Delta G + \frac{\del P_{IF}}{\del H}\Delta H = 0.
\end{equation}
Using Equations \ref{pif} and \ref{dpif}, we find that the necessary fractional change in IF amplification is related to the fractional change in mixer conversion gain by
\begin{equation}
\label{dadg}
\frac{\Delta H}{H} = -\left(1-\frac{P_H}{P_{IF}}\right)\frac{\Delta G}{G}.
\end{equation}
where we have introduced $P_{H} \equiv k_B B H T_H$, the output noise power of the HEMT amplifier.

If we further assume that the conversion gain of the mixer is only a function of its physical temperature, $T_{phy}$, we can expand $G(T_{phy})$ in a Taylor series around its nominal operating temperature, $T_0$, so that
\begin{equation}
\label{gt}
G(T_{phy}) = G(T_0) + \left.\frac{dG}{dT_{phy}}\right|_{T_0}\!\!\!\Delta T + \frac{1}{2!} \left.\frac{d^2G}{dT_{phy}^2}\right|_{T_0} \!\!\!\left(\Delta T\right)^2 + \ldots
\end{equation}
where the observed physical temperature fluctuations, $\Delta T$, are small compared to the nominal operating temperature (generally $\Delta T/T_0 \simlt 5\%$).  We then linearize Equation \ref{gt} by ignoring terms of order $\left(\Delta T\right)^2$ and higher to obtain 
\begin{equation}
\label{dg}
\Delta G \equiv G\left(T_{phy}\right) - G\left(T_0\right) \approx  g_T\Delta T.
\end{equation}
where $g_T$ is equal to the first derivative of the curve of mixer conversion gain versus physical temperature, evaluated at $T_0$.  Over a broad temperature range, $g_T$ is well approximated by the derivative of the Fermi function \cite{refs:Kooi}.  Over temperatures relevant to this work, $G(T_{phy})$ is approximately quadratic so $g_T$ varies linearly with physical temperature as confirmed by our measurements described in Section \ref{mixerConversionGain}.

In our system, we alter the IF gain by changing the bias current of the final stage of the low noise 3 stage HEMT amplifier immediately following the mixer.  
The HEMT gain, $H$, is a linear function of its bias current, $I_B$, as shown in Section \ref{hemtGainVsCurrentSection}.  Thus 
\be
\label{aVsI}
H(I_B) \approx H(I_0) + h_I \Delta I
\ee
where $h_I$ is the slope of the HEMT gain versus bias current curve evaluated at current $I_0$ and $\Delta I$ is a small change in the HEMT bias current.

With the assumption that the HEMT gain depends only on its bias current, in other words, by ignoring its dependence on physical temperature (in practice, temperature induced HEMT amplifier gain fluctuations are much smaller than mixer gain variations, as shown in Figure \ref{mixerGainVsTemp}),
we have
\be
\label{da}
\Delta H \equiv H\left(I_B\right) - H\left(I_0\right) \approx h_I \Delta I
\ee

Thus, from Equations \ref{dadg}, \ref{dg} and \ref{da}, we see that the necessary change in HEMT bias current for a given change in the physical temperature of the mixer is given by
\be
\Delta I = -\frac{g_T}{h_I}\frac{H}{G}\left(1-\frac{P_H}{P_{IF}}\right)\Delta T
\ee

In order to change the HEMT third stage bias current, we generate a voltage with a digital to analog converter, $V_{DAC}$, and feed it through a series resistor, $R$.
Therefore the required DAC voltage is 
\begin{eqnarray}
\label{servoEqn}
V_{DAC} &=& R \Delta I \\
&=& -\left[R\frac{g_T}{h_I}\frac{H}{G}\left(1-\frac{P_H}{P_{IF}}\right)\right]\Delta T \\
\label{servoEqn2}
&=& -f\Delta T
\end{eqnarray}
where $f$ is the proportional gain of the open-loop servo system, an experimentally determined value, with units of V K$^{-1}$, that relates changes in the physical temperature of the mixer to the DAC voltage necessary to stabilize the receiver.

Because $f$ is not known ahead of time and it changes with operating conditions, we have a training period of 5-10 minutes during which we place a fixed temperature load in the receiver beam and select a value of $f$ to minimize $dP_{IF}/dT_{phy}$.  Then the calibration load is removed from the beam and the value of $f$ is maintained during the astronomical observation.  We routinely achieve a stability of a part in 6,000 during the training period, and the derived value of $f$ is typically valid over many hours.

\subsection{Minimizing Changes in Receiver Noise}
When we change the IF gain, we introduce variations in the receiver noise temperature.  
Receiver noise temperature fluctuations degrade the receiver sensitivity in the same manner as gain fluctuations \cite{refs:Tiuri}.
The magnitude of this effect depends on how far along the IF chain the gain fluctuation is introduced, with the largest effect occurring if the gain of the first stage is changed.  For $k$ stages of cascaded amplification each with noise temperature $T_{N_j}$ and gain $G_j$ the total noise temperature can be written
\be
T_N = T_{N_1}+\frac{T_{N_2}}{G_1}+\frac{T_{N_3}}{G_1G_2}+ \ldots + \frac{T_{N_k}}{G_1G_2\cdots G_{k-1}}. 
\ee


We choose to vary the gain of the third (final) stage of the low noise HEMT amplifier.  
For typical values of IF amplification ($\sim$50 dB) and noise temperatures (80 K at 230 GHz) at the SMA, we find that the fractional change in receiver noise temperature for a 1\% change in the gain of the HEMT amplifier's third stage is 3 x $10^{-5}$, which is near the fundamental radiometer noise limit ($1/\sqrt{B\tau}$) of 2 x $10^{-5}$ for a 1 second integration with the SMA's 2.5 GHz IF bandwidth, and sufficiently less than our targeted stability of $10^{-4}$.



\subsection{Preserving Signal Phase}
Changes in the bias current to the HEMT amplifier's third stage will also alter the phase of the IF signal.  The process of downconverting the sky frequency to the IF preserves the phase of the source.  In other words, any phase error introduced into the IF signal will appear as a phase error of equal magnitude in the source signal.  
In interferometry, phase errors from an antenna degrade the correlated source signal.  In Section \ref{ifphasesection}, we present experimental results showing that our servo system introduces only negligible phase errors into the IF signal.

\section{Experimental Setup}
We have made stability measurements using a 230 GHz SIS heterodyne receiver designed for the SMA \cite{refs:SMARX}.  A schematic of the receiver and the open-loop gain stabilization system is shown in Figure \ref{servoSchematic}.  The horn, mixer, isolator and 3 stage low noise HEMT IF amplifier with $\sim$30 dB gain are maintained at a nominal temperature of 4.2 K, and an additional gain of 20 dB is provided by an amplifier maintained at 20 K (not shown in Figure \ref{servoSchematic}).  
The receiver output, 2.5 GHz wide centered at 5 GHz, is measured with a Herotek DT-4080 tunnel diode that is temperature stabilized near room temperature to 1 part in $10^5$ (rms to mean ratio).  The rms stability of this continuum detector was measured at 1 part in 8,000 over 600 seconds with a 33 ms integration time.  This value is $\sim$15\% larger than the $1/\sqrt{B\tau}$ fundamental radiometer noise and sets a noise floor for our gain stabilization system.
The detected signal is digitized by an Analog Devices AD7716 22-bit, 4 channel sigma-delta analog to digital converter (ADC) at 230 Hz and averaged down to a $\sim$30 Hz sampling rate (33 millisecond integration time) in software.  The physical temperature of the mixer is measured with a Lakeshore DT-470 silicon diode and recorded simultaneously by the same ADC.  Temperature fluctuations, $\Delta T$, are computed by differencing each temperature measurement with the first temperature measurement of the time series.  The microprocessor then instructs the digital to analog converter (DAC) to output a voltage proportional to $\Delta T$, as described in Equation \ref{servoEqn2}.  The analog voltage from the DAC, $V_{DAC}$, is fed through the resistor, $R$, to generate an excess bias current for the third stage of the HEMT IF amplifier.  
This excess current is used to change the HEMT gain (see Section \ref{hemtGainVsCurrentSection}) to compensate for the physical temperature induced conversion gain fluctuations in the mixer.

During a servo system training period, the receiver observes a constant temperature source, for which we used a microwave absorber immersed in liquid nitrogen.  The proportional loop gain factor, $f$, is selected to minimize the correlation between the physical temperature of the mixer and the detected IF power.





\section{Measurements and Discussions}

\subsection{Mixer Conversion Gain}
\label{mixerConversionGain}
\begin{figure}
\centering
\includegraphics[angle=0, scale=0.4]{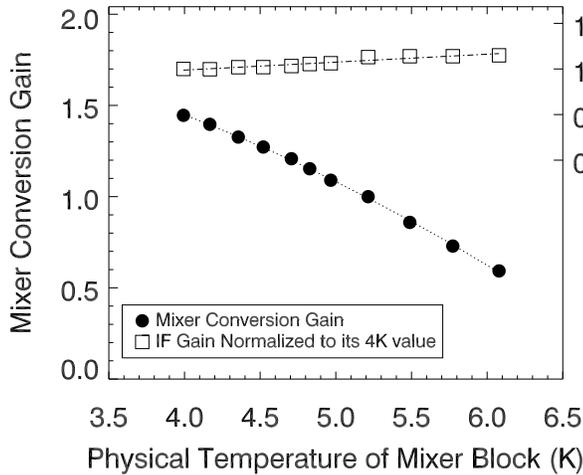}
\caption{The conversion gain of the 230 GHz SIS mixer and the normalized IF gain as a function of the mixer physical temperature.  The mixer data are well fit by the displayed quadratic $2.006+0.042*T_{phy}-0.045*T_{phy}^2$ over entire temperature range.  The slope of this curve, $g_T$, increases in magnitude from -0.322 K$^{-1}$ at 4 K to -0.504 K$^{-1}$ at 6 K.  The IF gain is a weaker function of temperature, with less than an 8\% change over the full temperature range.  It is well fit by the linear function $0.820+0.0435*T_{phy}$.  
Over a typical operating temperature swing of 50-100 mK, the IF gain would vary by 0.22-0.44\%.}
\label{mixerGainVsTemp}
\end{figure}

By adjusting the J-T valve on the cryocooler, we varied the physical temperature of the mixer from 4 K to 6 K.  From this data, we then determined the conversion gain of the mixer and the gain of the IF chain as functions of the physical temperature, using the technique described in \cite{refs:Woody}.  
We sampled this temperature range in steps of $\sim$0.2 K.  Figure \ref{mixerGainVsTemp} shows that over the entire 2 K temperature range, the data are well fit by a quadratic and that the 
gain of the mixing element is more sensitive to physical temperature fluctuations than is the IF gain.
At the SMA, the amplitude of cryocooler-induced temperature fluctuations is typically 50-200 mK.  Over such a small fractional temperature range (1-5\% of 4.2 K), the temperature dependence of the mixer conversion gain is well modeled by a straight line and the linear approximation to the Taylor series expansion of $G\left(T_{phy}\right)$ (Equation \ref{dg}) is sufficient.  
From the quadratic fit to $G(T_{phy})$, we compute that $g_T$, the derivative of the gain curve, increases in magnitude with temperature from -0.322 K$^{-1}$ at 4 K to -0.504 K$^{-1}$ at 6 K.

\subsection{HEMT Amplifier Gain}
\label{hemtGainVsCurrentSection}
\begin{figure}
\centering
\includegraphics[angle=0,scale=0.42]{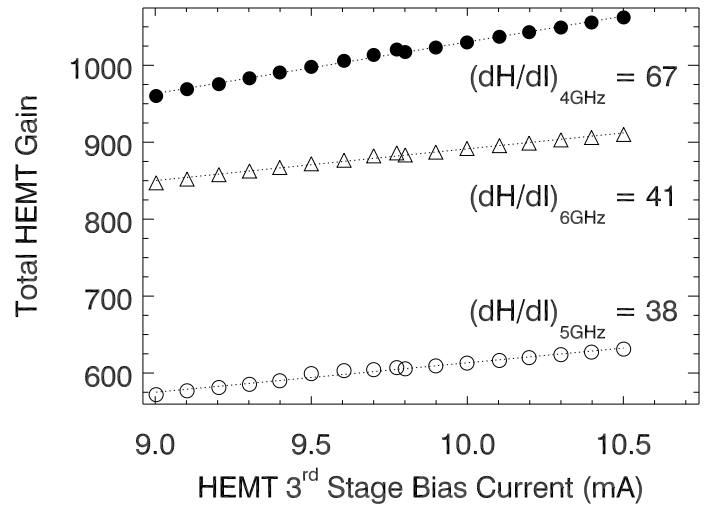}
\caption{The gain of the HEMT IF amplifier at room temperature as a function of its the third stage bias current at 4, 5 and 6 GHz, and corresponding linear fits.  The slope of the gain curve, $h_I$, is frequency dependent.  Nevertheless, we can achieve a receiver gain stability of 1 part in 6,000 using this technique.}
\label{hemtGainVsCurrent}
\end{figure}

The majority of the SMA receivers incorporate three stage HEMT amplifiers fabricated by the National Radio Astronomy Observatory (NRAO) with a total gain of $\sim$30 dB.
Using an 8510 HP network analyzer, the total HEMT gain at 4 GHz, 5 GHz and 6 GHz (IF band center and edges) was measured at room temperature while we varied the bias current of the third stage from 9.0 mA to 10.5 mA.  As shown in Figure \ref{hemtGainVsCurrent}, the gain is well fit by a straight line over the entire range of bias currents for all measured frequencies, so $h_I$ is nearly constant over the entire range of bias currents at a given frequency.  When the gain stabilization scheme is active, the bias current changes by $\simlt$0.1 mA and over this range, $h_I$ is sufficiently constant.  We note that $h_I$ has a frequency dependence that may ultimately limit the usefulness of this gain stabilization technique.  

\begin{figure}
\centering
\includegraphics[angle=0, scale=0.42]{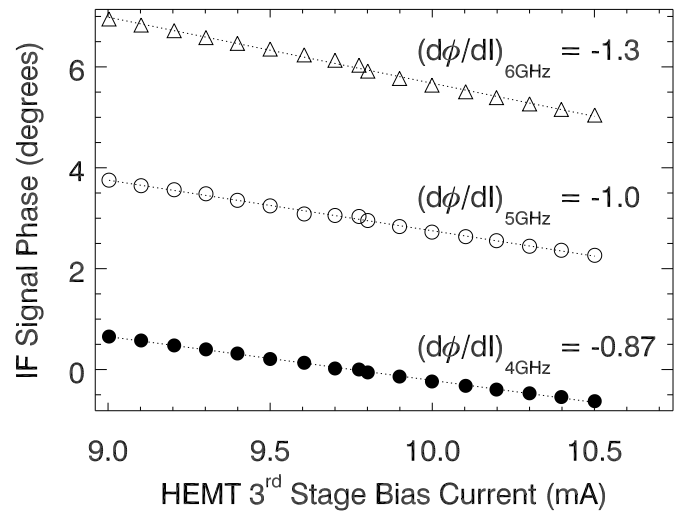}
\caption{The phase, $\phi$, of the IF signal out of the HEMT amplifier as a function of its the third stage bias current at 4, 5 and 6 GHz, and corresponding linear fits.  The phase data has been mean subtracted and offset for clarity.}
\label{hemtPhaseVsCurrent}
\end{figure}

\begin{figure*}
\centerline{\subfigure[Time Series Data]{
\includegraphics[scale=0.42, angle=0]{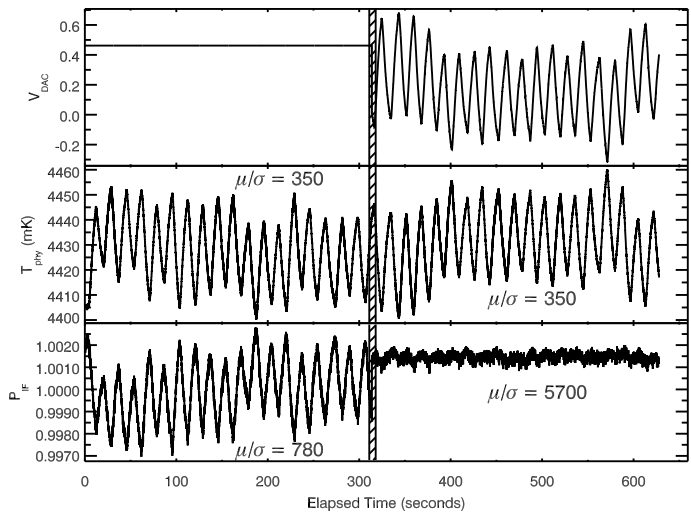}
\label{servoJTTime}}
\hfil
\subfigure[Correlation Between Continuum Power and Mixer Temperature]{
\includegraphics[scale=0.42, angle=0]{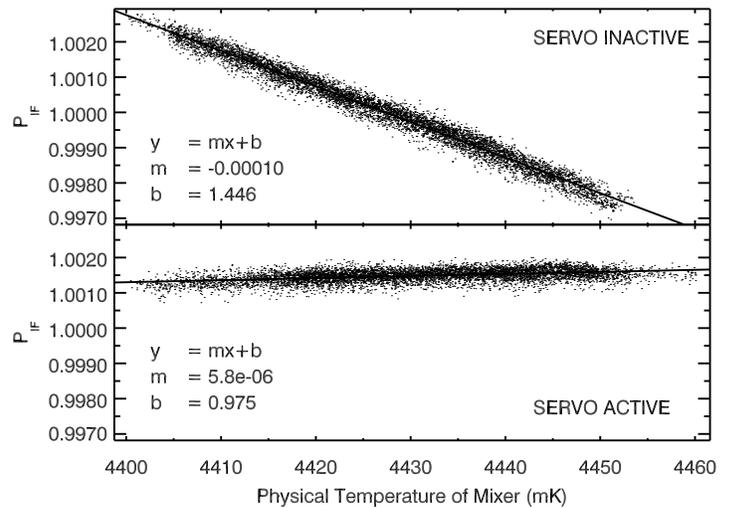}
\label{servoJTCorr}}}
\caption{{\bf At left:}  Time series data of the physical temperature of the mixer (middle plot) and the normalized detected continuum power (bottom plot).  The top plot presents the voltage, $V_{DAC}$, applied to the current bias circuitry of the third stage of the HEMT amplifier.  Here the J-T valve is continuously manually adjusted to introduce $\sim$50 mK peak to peak fluctuations in the physical temperature of the mixer, a stability typical of cryocoolers at the SMA.  For the first 311 seconds the servo system is inactive, so $V_{DAC}$ is constant.  There is a clear correlation between the physical temperature of the mixer and the detected IF power.  From 318 seconds onward, the gain stabilization scheme is active and there is a marked reduction in the amplitude of the IF power fluctuations.  The data in the shaded interval are omitted from the analysis.  During this time the servo system is in transition from inactive to active.  The mean to standard deviation stability parameters, $\mu/\sigma$, are shown for the mixer temperature and IF power in the relevant plot regions.  The gain stabilization servo system improves the receiver stability by a factor of $\sim$7.3 from 1 part in 780 to 1 part in 5,700 (rms to mean ratio).  {\bf At right:}  The correlation plots show that the activation of the gain stabilization servo system reduces the slope of the correlation between the IF power and the temperature of the mixer.  The residual correlation between the IF power and physical temperature of the mixer with the servo enabled indicates that we have slightly overestimated the servo loop gain ($f$ in Equation \ref{servoEqn}).}
\label{servoJT}
\end{figure*}
\begin{figure*}
\centerline{\subfigure[Time Series Data]{
\includegraphics[scale=0.42, angle=0]{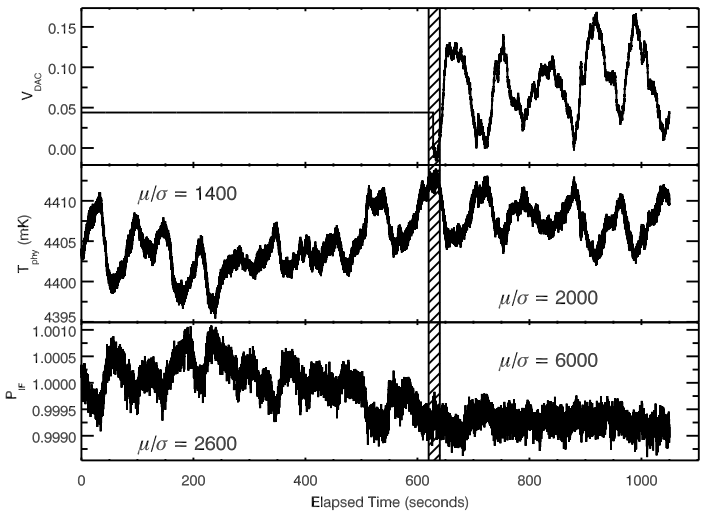}
\label{servoNoJTTime}}
\hfill
\subfigure[Correlation Between Continuum Power and Mixer Temperature]{
\includegraphics[scale=0.42, angle=0]{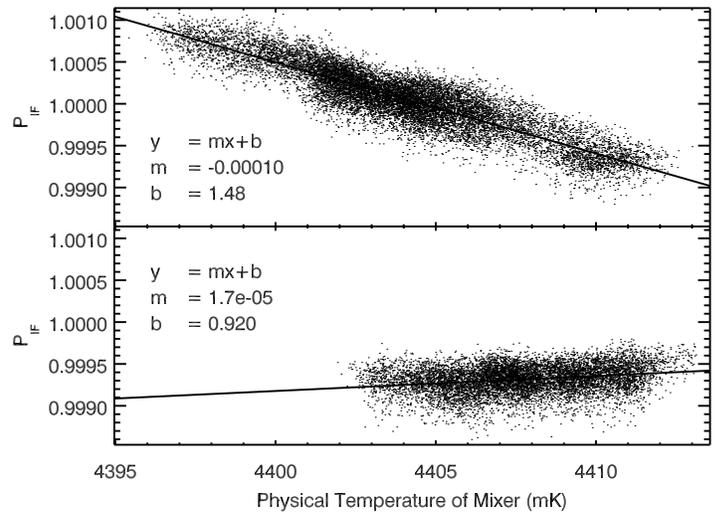}
\label{servoNoJTCorr}}}
\caption{Same as Figure \ref{servoJT}, but this time with a freely running cryocooler (i.e. the J-T Valve was not adjusted).  The temperature stability of the 4 K stage of this cryocooler is significantly better than a typical SMA cryocooler, with 4 K temperature fluctuations of only 3-5 mK rms.  Even with such a stable operating temperature, we improve the receiver stability by a factor of $\sim$2.4 with our gain stabilization servo system.}
\label{servoNoJT}
\end{figure*}

\subsection{IF Phase}
\label{ifphasesection}
At the same time that we measured the effect of bias current on the gain of the HEMT amplifier (see previous section), we also measured the effect of the bias current on the signal phase.  Figure \ref{hemtPhaseVsCurrent} shows that the phase changes linearly with the bias current and that the slope of the phase error versus bias current increases with frequency.  The maximal phase gradient ($d\phi/dI_B$) is -1.3 degree mA$^{-1}$.  
During normal servo operation, the bias current is changed by $\simlt$0.1 mA, and therefore the servo system will introduce $<<1$ degree of phase error in the IF signal.  Using data from \cite{refs:Masson}, we find that under median weather conditions on Mauna Kea on a 100 meter baseline, one expects 0.12 degrees of atmosphere induced phase fluctuations per GHz of observing frequency (27 degrees at 230 GHz).


\subsection{The Gain Stabilization System Performance}
Figure \ref{servoJTTime} shows the response of a 230 GHz receiver observing an absorber immersed in liquid nitrogen.  
The cryocooler in use during these tests had a very stable 4 K stage temperature compared to a typical SMA cryocooler.  To simulate temperature instabilities more typical of SMA cryocoolers, we manually adjusted the J-T valve of the cryocooler to induce periodic $\sim$50 mK peak to peak fluctuations in the physical temperature of the mixer.
Figure \ref{servoJTTime} shows the time variation of $V_{DAC}$, which generates the bias current correction for the third stage of the HEMT amplifier, $T_{phy}$, the physical temperature of the mixer and $P_{IF}$, the detected IF power.  $P_{IF}$ is normalized to its mean during the time when the gain stabilization system is inactive. 

During the first 311 seconds of the data set, $V_{DAC}$ is held constant so that the servo system is inactive.  The direct correlation (with negative proportional factor) between the physical temperature of the mixer and the receiver output is observed.  When the physical temperature of the mixer increases, its conversion gain decreases and the receiver output decreases.  
As seen in the bottom plot of Figure \ref{servoJTTime}, the resulting instrumental gain fluctuations limit the receiver stability to 1 part in 780 (rms to mean ratio).

From 318 seconds onward, when the gain stabilization system is activated, $V_{DAC}$ is set in proportion to fluctuations in the physical temperature of the mixer.  The resulting gain change in the third stage of the HEMT amplifier reduces the fluctuations in IF power caused by mixer conversion gain fluctuations, and the IF power stability improves by a factor of $\sim7.3$ from 1 part in 780 to 1 part in 5,700.

Figure \ref{servoJTCorr} demonstrates that the IF power depends linearly on the physical temperature of the mixer without the servo loop.  
A linear fit to the correlation between the IF power and physical temperature of the mixer is shown and it has a slope of $-10^{-4}$ mK$^{-1}$ with a negative sign indicating that an increase in the physical temperature of the mixer corresponds to a decrease in receiver output.  

Figure \ref{servoJTCorr} also demonstrates that the gain stabilization servo system decreases the slope of the correlation between the IF power and the physical temperature of the mixer.  The data for 318 seconds onward are plotted along with the best-fit line with slope 5.8 x 10$^{-6}$ mK$^{-1}$.  The positive slope indicates that the servo system has slightly over-corrected for the mixer conversion gain fluctuations.  Nevertheless, the receiver output power has been significantly decorrelated from the temperature of the mixer.



In Figure \ref{servoNoJT} we present the results of the same tests performed with a freely running cryocooler (i.e. with no J-T adjustment).
The order of the plots are the same as in Figure \ref{servoJT}.
Fluctuations in the physical temperature of the mixer are largely due to the G-M cycle of the cryocooler on about a one second time scale, and a larger amplitude, $\sim$60 second semi-periodic fluctuation of unknown origin.
In addition, there is a low frequency drift ($\sim$5 minute period) in the physical temperature of the mixer.
Here, the intrinsic temperature stability of the cryostat and thus the IF power is very good even before activating the servo system.  Still, the gain stabilization servo loop improves the receiver stability from 1 part in 2,600 to 1 part in 6,000.  
Figure \ref{servoNoJTCorr} shows that the proportional coefficient between IF power and mixer temperature has been reduced to 1.7 x 10$^{-5}$ mK$^{-1}$.

The effective temperature fluctuations, $\Delta \tilde{T}$, that would generate the observed residual IF power fluctuations in the absence of any receiver stabilization can be written
\be
\label{teff}
\Delta \tilde{T} = \left|\left(\frac{dP}{dT}\right)^{-1}\Delta P\right|
\ee
where $dP/dT$ is the slope of the IF power versus mixer temperature plot and $\Delta P$ is the residual fluctuation in continuum power when the servo is active.  From the top part of Figure \ref{servoJTCorr} we find that $dP/dT = -10^{-4}$ mK$^{-1}$.  The bottom part of Figure \ref{servoJTCorr} shows that $\Delta P$ = 0.0013 peak to peak or $\Delta P$ = 1.7 x 10$^{-4}$ rms (in normalized units).  Therefore the gain stabilization servo system achieves an effective peak to peak temperature stability of 
\be
\Delta \tilde{T} = 13 \mbox{ mK}
\ee
and an effective rms temperature stability of 
\be
\sigma_{\tilde{T}} = 1.7 \mbox{ mK.}
\ee

\subsection{Limitations of This Technique}
This IF power stabilization servo system responds to changes in the physical temperature of the mixer.  In practice, variations in other parameters, such as LO power, IF amplifier gain and mixer bias voltage will affect the IF power level.  In our receiver these fluctuations are small enough to allow receiver gain fluctuations of 1.6 parts in $10^{4}$ over many minutes.  Our continuum power stability limit of 1 part in 6,000 could be improved by identifying the next leading source of IF power fluctuations and incorporating it into the servo loop and by reducing the noise floor of our continuum detector.


The proportional loop gain factor, $f$, (see Equation \ref{servoEqn}) will change with time and operating conditions (e.g. the mean physical temperature of the mixer could drift from its nominal 4.2 K value), and so the servo loop must be calibrated periodically.  The time scale for this calibration is under investigation, but a single proportional factor typically provides 1 part in 6,000 receiver output stability over a period of several hours.  
In addition, $f$ depends on the total IF power, $P_{IF}$.  At present, we use either an ambient load or a liquid nitrogen load for calibration.  The median sky opacity at 225 GHz on Mauna Kea is 0.1, corresponding to a sky brightness temperature of 28 K.  To account for this, a procedure to determine $f$ should be incorporated into the standard calibration routine that is run prior to astronomical observations.

\section{Conclusion}
We have developed a system that stabilizes the gain of a 230 GHz SIS heterodyne receiver to 1 part in 6,000 (rms to mean).  With the use of an open-loop proportional servo system, we monitor the physical temperature of the mixer block and adjust the gain of the third and final stage of a low noise HEMT IF amplifier in order to compensate for the subsequent variations in the conversion gain of the mixer.
We have shown that the conversion gain of the mixer varies linearly over the range of typical cryocooler-induced physical temperature fluctuations, and that the gain of the HEMT amplifier varies linearly with its third stage bias current.  Using our gain stabilization system, we routinely achieve a total receiver gain stability of 1 part in 6,000 (rms to mean ratio) for a 0.033 second integration time over 10 minutes, which corresponds to an effective rms temperature fluctuation of 1.7 mK.  
In comparison, the typical fluctuations in the physical temperature of the mixer at the SMA are on the order of 50-100 mK, and the corresponding power fluctuations are 1 part in several hundred.  
Therefore, our gain stabilization servo system can provide more than an order of magnitude improvement over a typical unstabilized system.  Our technique introduces a negligible level of instrumental phase error into the source signal.  
This gain stability system can be implemented on any heterodyne receiver system in which the physical temperature of the mixer can be monitored and the IF amplification can be adjusted in real time.



\bibliographystyle{IEEEtran}
\bibliography{refs}

\end{document}